\documentclass{elsart}
\usepackage{epsfig}

\newcommand{\centps}[2]{
	\begin{center}
		\epsfig{file=#1,width=#2mm}
	\end{center}
}

\newcommand{\gdot}{\dot{\gamma}}
\newcommand{\mbar}{s_0}
\newcommand{\tgel}{t_{\rm gel}}

\begin{document}


\begin{frontmatter}

\title{Shear-Induced Clustering in a Simple Driven Diffusive Model}
\author{O.J.~O'Loan, M.R.~Evans and M.E.~Cates}
\address{Department of Physics and Astronomy, University of Edinburgh,
      	Mayfield Road, Edinburgh EH9 3JZ, U.K.}


\begin{abstract}
We study a simple lattice model of shear-induced clustering in two dimensions
in which clusters of particles aggregate under an imposed shear flow and
fragment stochastically. Two non-equilibrium steady states are identified: an
unjammed state and a jammed state characterised by a system-spanning cluster. A
discontinuous jamming transition with strong hysteresis occurs as the shear
rate is increased or fragmentation rate decreased. We study the kinetics of
jamming and measure power law cluster size distributions. We also consider some
general simulation issues including the role of Galilean invariance.
\end{abstract}

\end{frontmatter}


\section{Introduction}			\label{sec:Intro}

When subjected to shear flow, many complex fluids exhibit shear thickening
behaviour where viscosity increases with shear rate. Shear thickening is often
discontinuous and may be accompanied by hysteresis and structural changes.
Systems where such behaviour has been observed include concentrated colloidal
suspensions \cite{Colloids} and rodlike micelle solutions \cite{Micelle}.

In the case of colloidal suspensions, it has been suggested that shear
thickening is the result of the formation of hydrodynamic clusters
\cite{BB}. These clusters comprise large numbers of particles bound together
by short-range lubrication forces and form along the compression axis of the
shear. Indeed a detailed analysis of the dynamics of the aggregation of such
rotating clusters, in the absence of Brownian motion, has recently been
performed \cite{FMB} and suggests that the discontinuous jump in viscosity is
due to a jamming transition where, essentially, a log-jam of clusters occurs.

Jamming transitions have also been studied in an apparently different context,
that of driven diffusive systems \cite{SZ}. In one dimensional models, jams
have been induced by the presence of defects or disorder \cite{Krug} or by a
non-conserved quantity mediating an effective long-range interaction
\cite{OEC}. In two-dimensional ($2d$) driven lattice gases, gridlock effects
have been studied \cite{BML,SHZ}.

Recently, Santra and Herrmann introduced a simple $2d$ lattice model of
gelation under shear \cite{SH}, which provides a link between colloidal systems
and driven lattice gases. They showed that within the model one finds dynamical
scaling as clusters of particles aggregate irreversibly, and that the clusters
are aligned roughly along the compression axis of the shear. The model has been
extended to take into account the effect of rotation of the clusters \cite{KB}.

In the present work, we extend the model of \cite{SH} in a different direction
by allowing clusters to fragment stochastically. This models thermal
fluctuations in a simple manner and gives rise to the possibility of
non-trivial steady states, in contrast to the absorbing state consisting of a
single cluster which is always attained with the models of \cite{SH} and
\cite{KB}. Indeed, as we shall show below, we find two types of steady state --
{\it unjammed}, where clusters have some typical size much smaller than the
transverse size of the system, and {\it jammed}, in which there are clusters
which span the system. We provide evidence of a discontinuous transition
between the two states, with hysteresis, as the fragmentation rate or the shear
rate is varied. We study the kinetics of the jamming process and find evidence
for dynamical scaling for both zero and non-zero fragmentation rate. Although
there is an intriguing similarity with the colloidal jamming problem studied in
\cite{FMB}, in practice the transitions we observe arise at very low
concentration. In this respect the behaviour is more like that of the micellar
problem of \cite{Micelle}.

The paper is organised as follows. In section \ref{sec:Model} we define the
model that we study and simulation results are presented in section
\ref{sec:Results}. In section \ref{sec:Issues} more general issues concerning 
the simulations are raised. We conclude in section \ref{sec:Conclusion}.


\section{Description of the Model}		\label{sec:Model}

\begin{figure}[th]
	\centps{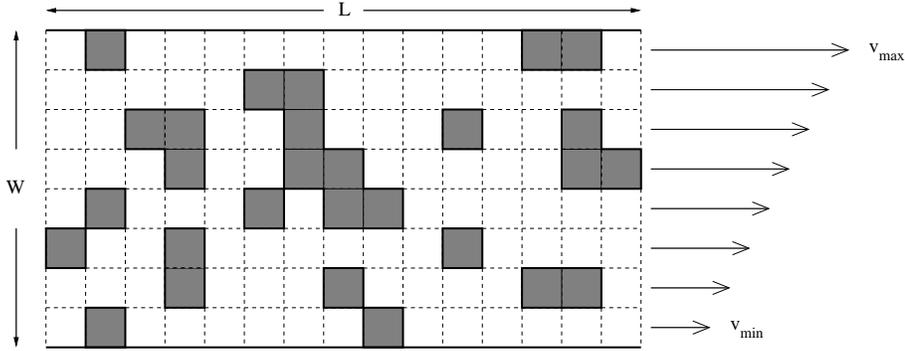}{120}
	\caption{Example lattice configuration. The shear flow is represented by
	the arrows.}
	\label{fig:Schematic}
\end{figure}

The model is defined on a $2d$ lattice of length $L$ and width $W$, illustrated
by an example configuration in figure \ref{fig:Schematic}. Each lattice site is
either occupied by a particle or empty. The boundary perpendicular to the
direction of flow is periodic but no particles may pass through the boundaries
parallel to the direction of flow -- they are ``hard walls''. A simple shear
flow is imposed by assigning a velocity $v(y)$ to each ``lane'' in the
direction of flow given by
\begin{equation}
	v(y) = v(0) + \gdot y
\end{equation}
where $y$ runs from $0$ to $W-1$. Each particle in lane $y$ is said to have
velocity $v(y)$. Hence, the shear rate is
\begin{equation}
	\gdot = \frac{v_{\rm max}-v_{\rm min}}{W-1}
\end{equation}
where $v_{\rm min} = v(0)$ and $v_{\rm max} = v(W-1)$. We consider only
non-negative velocities from now on. There are $N$ particles in the system and
the concentration is
\begin{equation}
	c = \frac{N}{WL}.
\end{equation}

We define a cluster to be a set of particles connected successively by
nearest-neighbour bonds on the lattice. Therefore, the nearest neighbour sites
belonging to an individual particle are either empty or occupied by particles
in the same cluster. The velocity $u$ of a cluster is defined as the mean of
the velocities of its constituent particles. The proximity of a wall does not
affect the velocity of a cluster so that one may think of the particles
``slipping'' relative to the walls. (In contrast, in order to set up a shear
flow in a fluid by moving one or both of the walls, a no-slip condition between
the {\it fluid\/} and the walls is required.)

A (continuous) ``counter'' variable $q$ is associated with each cluster. In
each simulation time-step, three steps (flow, cluster merging and fragmentation)
are performed in the following sequence:
\begin{description}
	\item[Flow] The counter $q$ for each cluster is incremented by $u$, the
	cluster velocity. If $q \ge 1$ for any cluster, all the particles in that
	cluster move one site forward ({\it i.e.} in the direction of flow) and $q
	\rightarrow q - 1$. If $q < 1$, no action is taken.
	\item[Cluster merging] At the end of the previous step, some clusters may
	have become connected and these clusters are now merged. In merging two
	clusters, the resulting single cluster is assigned a counter which is the
	weighted average of the counters of the two constituent clusters
	labelled $i$ and $j$:
	\begin{equation}
		q_{i+j} = \frac{n_i q_i + n_j q_j}{n_i + n_j}
	\end{equation}
	where $n_i$ is the number of particles in cluster $i$. We have checked that
	this ensures that the average velocity of clusters in lane $y$ is indeed 
	$v(y)$.
	\item[Fragmentation] This random sequential step allows single particles to
	leave clusters, possibly resulting in the break up of clusters. The
	following update rules are repeated $N$ times:
	\begin{enumerate} 
		\item A particle is chosen at random.
		\item A direction for the particle to move is chosen at random from the
		four possibilities.
		\item If the particle is not blocked by another particle in the chosen
		direction, it moves one site in that direction with probability 
		$T$, where $T$ may be thought of as the ``temperature'' of the system.
		\item The moved particle becomes a single-particle cluster with counter
		variable equal to that of the cluster it previously belonged to. If it
		adjoins any other clusters, this single-particle cluster is then merged
		as detailed previously.
	\end{enumerate}
	The fragmentation update of one particle may result in one of five
	outcomes: no change, a particle leaving its cluster, a particle moving
	to a different location in the same cluster, movement of a single-particle
	cluster or a single-particle cluster joining another cluster.
\end{description}

The three physical parameters present in the model are concentration $c$, shear
rate $\gdot$ and ``temperature'' $T$. 

Simple shear flow is Galilean invariant, by which we mean that adding a
constant velocity at every point in the system has no effect on the physics.
It is therefore desirable that any model of simple shear is as close as
possible to being Galilean invariant. We discuss this issue in section
\ref{sec:Issues} where we show that lattice effects break Galilean invariance
in the model. In addition, we remark that the model is translationally
invariant only in the direction of flow. The fixed walls break translational
invariance normal to the direction of flow.


\section{Results}			\label{sec:Results}

\subsection{Steady State Behaviour}

\begin{figure}
	\centps{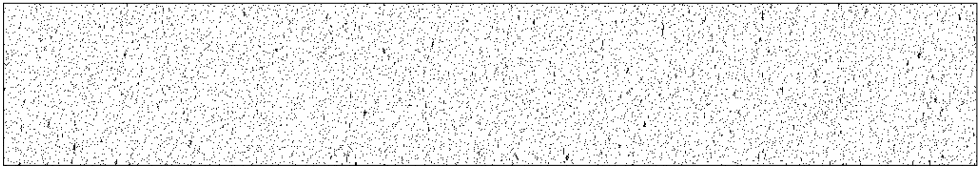}{130}
	\caption{Snapshot of an unjammed 1200$\times$200 system in the steady
	state. The simulation parameters were $c = 0.07$, $T = 0.0045$ and $\gdot =
	5/1990 \simeq 0.0025$. The flow is from left to right and the velocity
	increases from bottom to top.}
	\label{fig:Unjammed}
\end{figure}

From computer simulation of the model defined in section \ref{sec:Model}, we
have found two types of steady state. The first is an unjammed state, where
many small clusters are evenly distributed throughout the system. A snapshot of
a typical unjammed system is shown in figure \ref{fig:Unjammed}. There are
some clusters which are elongated perpendicular to the direction of flow but
the typical cluster size is very small.

\begin{figure}
	\centps{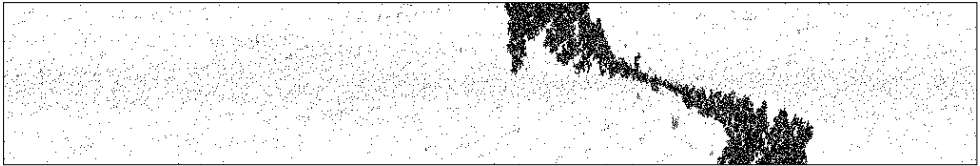}{130}
	\caption{Snapshot of a jammed 1200$\times$200 system in the steady
	state. The simulation parameters were the same as for figure
	\ref{fig:Unjammed} with the exception that here $\gdot = 7/1990 \simeq
	0.0035$.}
	\label{fig:Jammed}
\end{figure}

A second type of steady state exhibited by the model is characterised by the
presence of clusters which span the system; we refer to it as ``jammed'' state
(although the spanning cluster is not stationary). These {\em spanning
clusters} contain a finite fraction of the particles in the system. A snapshot
of a typical jammed system is shown in figure \ref{fig:Jammed}; the system is
strongly inhomogeneous in a number of respects. Since the spanning cluster is
moving with a velocity close to that of the particles in the centre of the
channel, the rate of aggregation is least in the centre and greatest at the
walls -- near the upper wall, small clusters move more quickly than the
spanning cluster and join it from behind whereas near the lower wall, small
clusters move more slowly than the spanning cluster and aggregate from the
front. Hence the spanning cluster is aligned at an angle to the direction of
flow and decreases in thickness from the edges to the centre of the channel.
Accordingly, the density in the rest of the system is greatest in the centre
and least at the edges. Some quite large clusters are visible close to the
spanning cluster -- these are probably pieces of the spanning cluster which
broke off some short time previously. The somewhat unusual, inhomogeneous
nature of the jammed steady state in the finite systems we study is primarily a
result of having hard wall boundaries parallel to the direction of flow rather
than any effect of the underlying square lattice.

\begin{figure}[thb]
	\centps{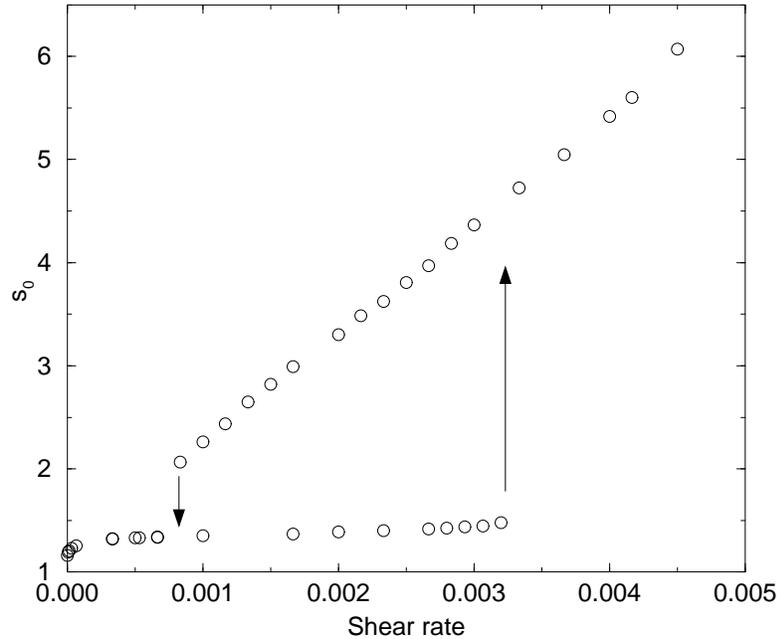}{120}
	\caption{Plot of $\mbar$ against $\gdot$ for a 900$\times$150 system
	with $c=0.07$ and $T=0.0045$. The direction of the hysteresis loop is shown
	by the arrows.}
	\label{fig:ng1}
\end{figure}

In order to discuss the behaviour of the model more quantitatively, we
introduce the cluster size distribution $n(m,t)$, the number of clusters
containing $m$ particles at time $t$. The moments $M_k(t)$ of the cluster size
distribution are given by
\begin{equation}
	M_k(t) = \sum_{m=1}^{\infty} m^k n(m,t)
\end{equation}
so that $M_1 = N$, the number of particles in the system and $M_0$ is the
number of clusters in the system. We define
\begin{equation}
	s_k(t) = \frac{M_{k+1}(t)}{M_k(t)}
	\label{eqn:def-s}
\end{equation}
as measures of typical cluster size \cite{VE1} for $k \ge 0$. In discussing
steady state properties of the system, we use the average cluster size $s_0$ as
our measure of typical cluster size since this quantity is not dominated by the
spanning clusters in jammed systems. In contrast, for $k > 0$, $s_k$ is of the
order of the size of a spanning cluster in the jammed state. We use $s_1$ when
discussing the kinetics of jamming in section \ref{sec:Kinetics}.

\begin{figure}[hbt]
	\centps{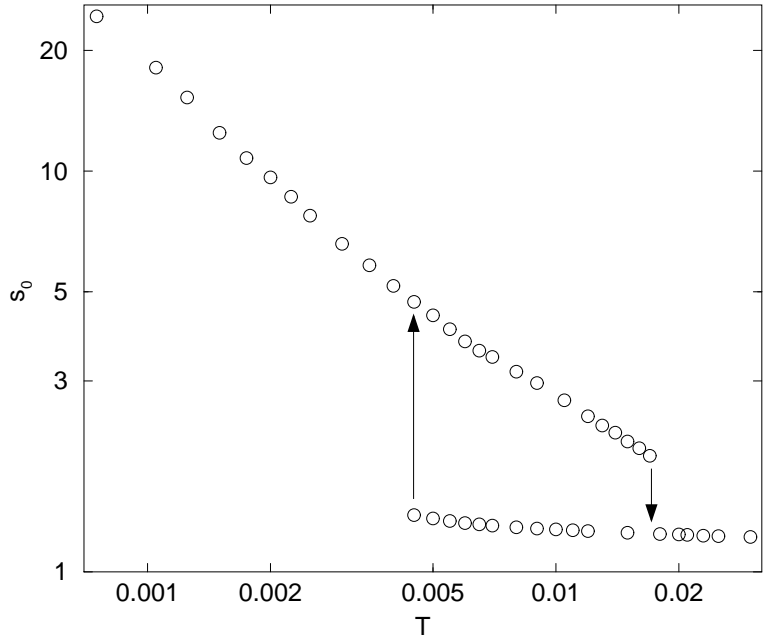}{120}
	\caption{Log-log plot of $\mbar$ against $T$ for a 900$\times$150 system
	with $c=0.06$ and $\gdot=6/1490 \simeq 0.004$. The direction of the
	hysteresis loop is shown by the arrows.}
	\label{fig:nT1}
\end{figure}

We now provide evidence of a strong discontinuous phase transition between
jammed and unjammed states as the parameters $T$ and $\gdot$ are varied.
Figure \ref{fig:ng1} shows a plot of $\mbar$ against $\gdot$. Strong hysteresis
is apparent, with the direction of the hysteresis loop shown by the arrows. As
$\gdot$ is increased from zero, $\mbar$ increases smoothly at first but jumps
discontinuously to a larger value, corresponding to a jammed system. When
$\gdot$ is then decreased, the jammed state remains stable (for at least $2
\times 10^5$ time-steps) for a range of $\gdot$ values in which an unjammed
state is also stable (for at least $10^5$ time-steps). The system becomes
unjammed again for sufficiently small $\gdot$ ({\it i.e.\/} an initially
jammed configuration becomes unstable). One can also see from figure
\ref{fig:ng1} that, for jammed systems, $\mbar$ increases roughly linearly 
with increasing shear rate.

Similar behaviour is found when $T$ is varied with $\gdot$ and $c$ held fixed.
Figure \ref{fig:nT1} shows a log-log plot of $\mbar$ against $T$. As in figure
\ref{fig:ng1}, strong hysteresis is apparent. There is a large range of values
of $T$ for which both unjammed and jammed states are stable (on a time-scale of
at least $2 \times 10^5$ time-steps for jammed systems and at least $10^5$
time-steps for unjammed systems.)

\begin{figure}
	\centps{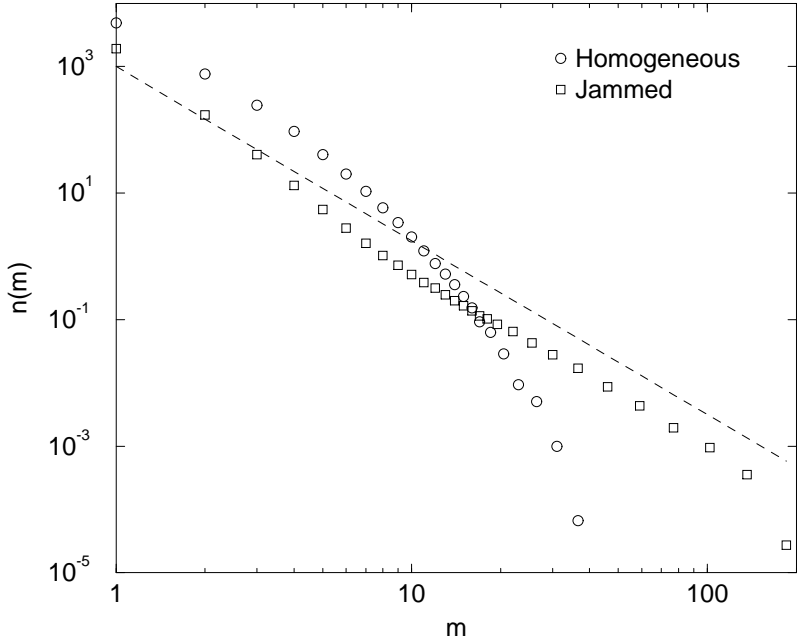}{120}
	\caption{Log-log plot of the steady state cluster size distribution for
	both unjammed and jammed systems with $T = 0.006$. The system size is
	900$\times$150, $c=0.06$ and $\gdot = 6/1490 \simeq 0.004$. The dashed line
	is proportional to $m^{-2.75}$ and is shown to guide the eye.}
	\label{fig:Dist-SS}
\end{figure}

Figure \ref{fig:Dist-SS} shows a log-log plot of the steady state cluster size
distribution $n(m)$ for both unjammed and jammed systems at the same value of
$T = 0.006$. In the homogenous system, $n(m)$ decays very rapidly and the large
$m$ behaviour appears to be exponential (it decays faster than any power
law). In the jammed system, the size distribution of small (non-spanning)
clusters only is shown; the distribution as a whole is bimodal, with the second
peak (not shown) corresponding to the spanning cluster. The figure suggests
that for the small clusters, $n(m)$ decays as a power of $m$ (for $m$ greater
than about $10$) with an exponent close to $-2.75$. However, $n(m)$ decays
faster than a power law for $m$ greater than about $100$. It is not clear
whether this is a finite system-size effect or the true asymptotic
behaviour. For jammed systems, the situation is further complicated by the fact
that the cluster size distribution is spatially inhomogeneous (see figure
\ref{fig:Jammed}).

\begin{figure}
	\centps{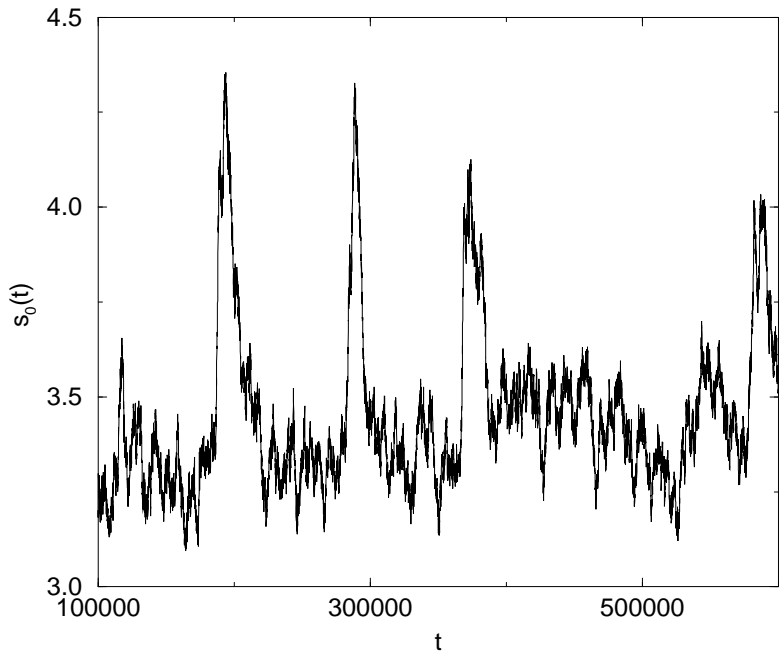}{120}
	\caption{Average cluster size $\mbar$ against time for a
	900$\times$150 system with $T=0.007$, $c=0.06$ and $\gdot=6/1490 \simeq
	0.004$.}
	\label{fig:mVt}
\end{figure}

We have found two different regimes of behaviour in the jammed state as $T$ is
varied. Figure \ref{fig:mVt} shows the variation of $\mbar$ with time in the
(jammed) steady state with $T=0.007$ (the other parameters are the same as for
figure \ref{fig:nT1}). The average cluster size $\mbar$ appears to fluctuate
about a value close to $3.25$ for some time before increasing suddenly to more
than $4.0$; it then gradually falls back to about $3.25$. We have found that
this behaviour is due to splitting and reforming of the spanning cluster. We
have already seen in figure \ref{fig:Jammed} that a spanning cluster is
narrowest at its centre. If $T$ is sufficiently small, the spanning cluster
does not split. However, when $T$ is sufficiently large the spanning cluster
may split in two, whereupon the two resulting clusters separate. At this point,
the two large clusters grow by aggregating with smaller clusters before meeting
again and reforming the spanning cluster (due to the periodic boundary in the
flow direction). Upon reforming, the spanning cluster is larger (it contains
more particles) than when it split; this is because the two large clusters
formed as a result of the split grow significantly before recombining. This
explains the sudden rises in average cluster size seen in figure \ref{fig:mVt}.
The newly reformed spanning cluster gradually loses particles before splitting
once more. The steady state is cyclic -- a spanning cluster is stable only in
the sense that once it splits, it always reforms. Close examination of figure
\ref{fig:nT1} reveals a kink in the jammed branch of $s_0(T)$ around
$T=0.007$. This is caused by the change in behaviour from a non-cyclic steady
state to a cyclic one. If $T$ is further increased beyond about $0.018$, the
two large clusters formed when a spanning cluster splits may become unstable
and break up (due to fragmentation dominating shear-induced aggregation). An
unjammed system then results and there is no longer a jammed steady state. It
is not clear to what extent the splitting of spanning clusters is affected by
system size.

\subsection{Jamming Kinetics}	\label{sec:Kinetics}

\begin{figure}
	\centps{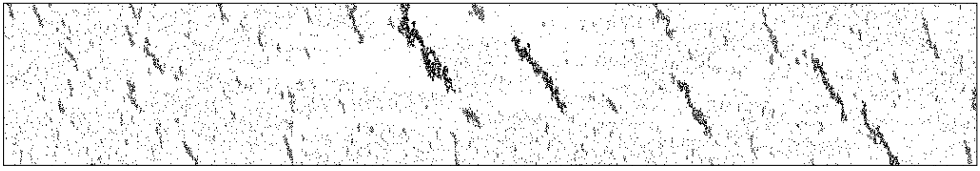}{130}
	\caption{Snapshot of a 1200$\times$200 system undergoing jamming at
	$t=7000$. The particles were positioned at random at $t=0$. The simulation
	parameters were $c = 0.06$, $T = 0.002$ and $\gdot = 8/1990 \simeq 0.004$.}
	\label{fig:Jamming}
\end{figure}

\begin{figure}
	\centps{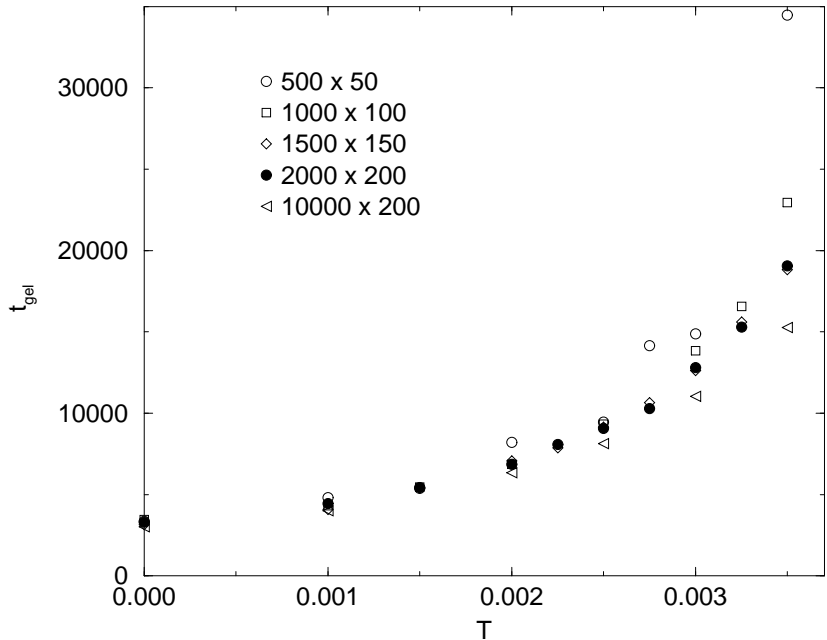}{120}
	\caption{Plot of $\tgel$ against $T$ for various system sizes. The
	simulations had $c=0.06$ and $\gdot=8/1990 \simeq 0.004$. The particles
	were positioned at random at $t=0$. We averaged over betwen 10 and 20
	independent runs.}
	\label{fig:tgel}
\end{figure}

The above concludes our discussion of the steady state properties of the model;
we now consider the approach to the steady state when the system spontaneously
jams from an initial condition of particles positioned at random. Figure
\ref{fig:Jamming} shows a snapshot of a system in the process of jamming 7000 
time-steps after the particles were positioned at random. Clusters with a large
range of sizes are visible.

We define the gel time $\tgel$ as the time at which a spanning cluster first
appears in the system. Santra and Herrmann \cite{SH} found (by extrapolation)
that, for $T=0$, $\tgel$ is finite for an infinite system. Figure
\ref{fig:tgel} shows a plot of $\tgel$ against $T$ for different system
sizes. It appears that as system size is increased for the fixed aspect ratio
$L/W=10$, $\tgel(T)$ converges. A fit to the data for the $2000\times200$
system indicated that $\tgel$ increases more quickly than $\exp(T)$ and it is
possible that it diverges for some finite value of $T$; this is consistent with
the presence of a phase transition at finite $T$. However, due to the large
amount of computational effort required to obtain $\tgel(T)$, we have been
unable to confirm this unambiguously.  Also, figure \ref{fig:tgel} shows that
$\tgel$ depends on the aspect ratio of the system since $\tgel$ for the
$10000\times200$ system is consistently smaller than for the $2000\times200$
system.

\begin{figure}
	\centps{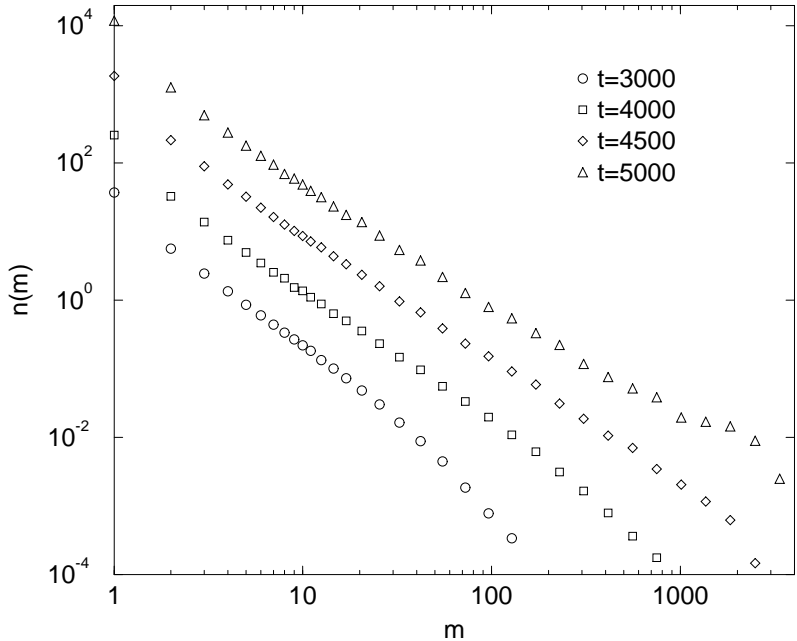}{120}
	\caption{Log-log plots of $n(m)$ at different times for $T= 0.001$. The
	system size was 10000$\times$200 and $c=0.06$, $\gdot=8/1990 \simeq
	0.004$. The particles were positioned at random at $t=0$. We averaged over
	12 independent runs. The data has been shifted vertically for clarity.}
	\label{fig:TimeDist}
\end{figure}

In a gelling system, we expect the pre-gelation cluster size distribution to
have the scaling form \cite{VE1,VE2}
\begin{equation}
	n(m,t) \sim m^{-\tau} f \left( \frac{m}{s_1(t)} \right)
	\label{eqn:scaling}
\end{equation}
for large cluster sizes, with the typical cluster size $s_1(t)$, given by
(\ref{eqn:def-s}), diverging as $t \to \tgel$. Since the number of particles
$N$ is conserved, we must have $\tau > 2$. The above scaling {\it ansatz\/} is
expected to apply for an infinite system. In our finite systems, we find
evidence of such scaling for $T \ge 0$ although finite-size effects are large.

Figure \ref{fig:TimeDist} shows a log-log plot of $n(m)$ for $T = 0.001$ at
four different times (note that the curves have been shifted vertically for
clarity). Power law behaviour appears gradually and is present over the largest
range of cluster sizes at $t^* \simeq 4500$. We define $t^*$ to be the time at
which we see power law behaviour of $n(m)$ over the maximum range of $m$. By
$t=5000$, spanning clusters have formed and a ``bump'' has begun to form in the
tail of the distribution. By fitting a power law to the $n(m)$ for large $m$ at
$t=t^*$, we find the exponent $\tau \simeq 1.9$. Since for an infinite system
$\tau$ must be greater than 2, we conclude that finite-size effects are
strong. Indeed, this is not surprising since we find, for $T \le 0.0025$, that
$t^* \simeq \tgel$ (as expected in gelling systems) so that there are clusters
with size of order the system size present at time $t^*$. However, for
$T=0.003$ and $T=0.0035$ we find that $t^*$ is somewhat smaller than $\tgel$,
an anomaly which we are unable to explain but which may be due to finite-size
effects.

\begin{figure}[htb]
	\centps{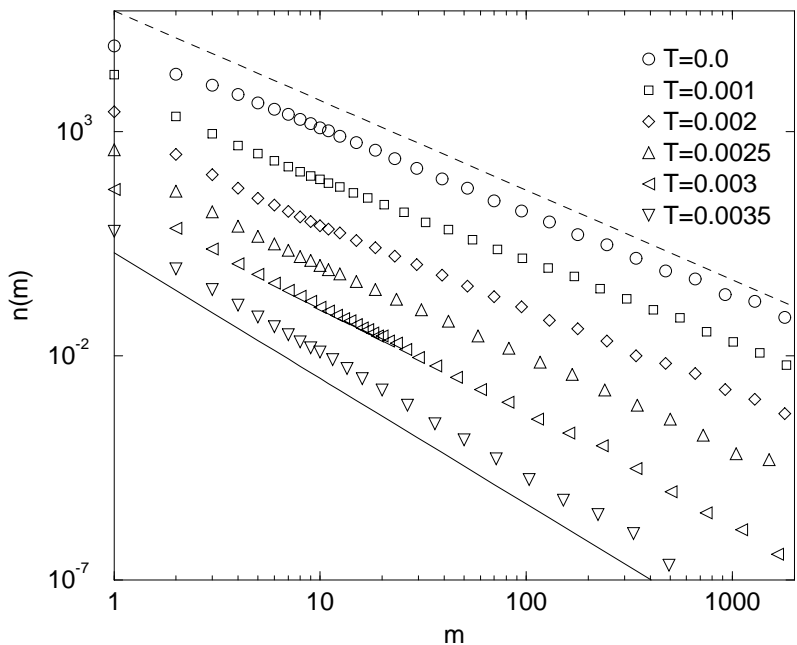}{120}
	\caption{Log-log plots of $n(m)$ for various values of $T$ after a time
	$t^*(T)$. The system size was 10000$\times$200 and $c=0.06$, $\gdot=8/1990
	\simeq 0.004$. The values of $t^*$ are 3400, 4500, 7000, 8500, 10000 and
	12000 (in order of increasing $T$). We averaged over 12 runs for each value
	of $T$. The data has been shifted vertically for clarity. The dashed line
	is proportional to $m^{-2}$ and the solid line is proportional to
	$m^{-2.8}$; they are shown to guide the eye.}
	\label{fig:SizeDist}
\end{figure}

Figure \ref{fig:SizeDist} shows log-log plots of $n(m)$ at $t^*$ for several
values of $T$. A power law cluster size distribution (for $m$ larger than about
10) is present for all values of $T$. The slopes of the power law regions
decrease with increasing $T$ (for $T > 0$), suggesting that the exponent $\tau$
varies significantly with $T$ and hence that it may be non-universal. Futher
careful study of finite-size effects would be required to clarify this point.


\section{Simulation Issues}			\label{sec:Issues}

The ``counter'' scheme \cite{SH} for performing cluster flow described in
section \ref{sec:Model} is not the only one that can be imagined. However, we
have found that it has important advantages over other schemes.

For example, a simpler way of implementing cluster motion in the direction of
flow is, at each time-step, to let each cluster move forward one site with
probability $v$, where $v$ is the velocity of the cluster. However, there are
seemingly insurmountable difficulties with this stochastic implementation of
cluster flow. To illustrate this point, let us consider how clusters aggregate
and grow. Aggregation occurs when one cluster catches up with a slower
neighbouring cluster. We wish the imposed shear to be the dominant factor in
the aggregation. However, if clusters move forward stochastically, as in the
scheme just described, fluctuations (in addition to the imposed velocity
gradient) will cause clusters to aggregate. For example, a cluster with
velocity $v_1$ may may move forward in several successive time-steps at the same
time as a cluster with velocity $v_2 > v_1$ does not move at all, possibly
resulting in the spurious aggregation of the clusters. Indeed, we have found
that when the stochastic implementation of cluster flow just described is used
(instead of the counter scheme described in section \ref{sec:Model}),
stochastic aggregation of clusters dominates aggregation due to the imposed
shear for practical parameter values.

A further advantage of the counter dynamics for flow is that it is much closer
to being Galilean invariant than a stochastic implementation of cluster flow.
This is because, for stochastic flow, the size of the fluctuations in the
motion of a cluster, and hence the rate of cluster-cluster aggregation, is
strongly dependent on the absolute magnitude of the velocity $v(y)$. It is
well-known that the variance of the displacement of an independent random
walker hopping in one direction with probability per time-step $v$ is
proportional to $v(1-v)$; we therefore expect the rate of cluster-cluster
aggregation for stochastic cluster flow to be greatest for $v = 0.5$ and
smallest for $v = 0$ and $v = 1$. We have confirmed that this is indeed the
case in simulations. This is a potentially serious problem in simulation of a
shear flow since $v$ varies across the width of the lattice.

\begin{figure}
	\centps{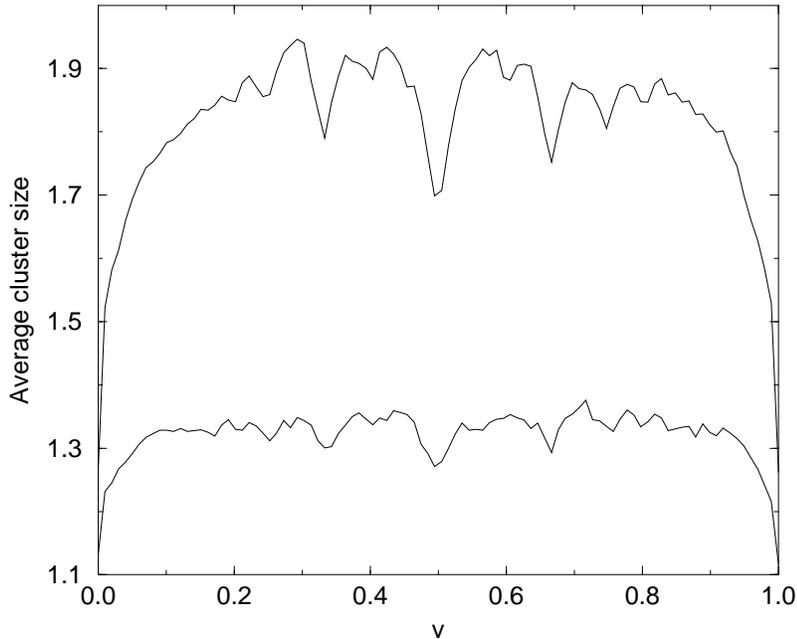}{120}
	\caption{Mean size of clusters having velocity $v$ in two unjammed
	systems (averaged over $10^5$ time-steps in the steady state). In both
	cases, the system size is $400 \times 100$ with $v_{\rm min} = 0$ and
	$v_{\rm max} = 1$. The top curve is for $c = 0.15$ and $T = 0.045$ while
	for the bottom curve, $c = 0.07$ and $T = 0.02$.}
	\label{fig:n-spatial}
\end{figure}

Since the counter implementation of cluster flow is deterministic, the problems
peculiar to stochastic cluster flow just discussed are circumvented. The
counter scheme is, however, not without its limitations. Figure
\ref{fig:n-spatial} shows the average cluster size of clusters whose centre of
mass lies in lane $y$ as a function of $v(y)$ for two unjammed systems with
different concentrations. The average cluster size is smallest for $v=0$ and
$v=1$ due to the presence of the fixed walls. However, the cluster size grows
quite slowly as one moves in from either wall and we have checked that the
reason for this is that $v$ is in the vicinity of either $0$ or $1$. Similarly,
a deep minimum in the average cluster size is evident around $v=1/2$, with less
pronounced minima also evident around $v=1/3, 2/3, 1/4, 3/4$, {\it etc\/}.
Clearly, the counter updating scheme has a systematic effect on the
cluster-cluster aggregation rate, breaking Galilean invariance by introducing a
velocity dependence of the cluster-cluster aggregation rate. Not surprisingly,
the effect diminishes as $c$ is decreased. In our simulations, we have
attempted to minimise the effect of the spatial inhomogeneity caused by the
updating scheme by working at low concentration and avoiding velocities less
than $0.1$ or greater than $0.9$.  For the concentration values we have used
($c \le 0.07$), it is clear that the shear is the dominant factor driving
cluster-cluster aggregation since the average cluster size varies significantly
with shear rate (see figure \ref{fig:ng1}).

It is clear that lattice models are not ideally suited to the simulation of a
Galilean invariant phenomenon such as shear-induced aggregation, although they
are a relatively simple and efficient approach to the problem. However, there
are physical situations where Galilean invariance is not respected, an example
being (multi-lane) traffic flow where the concept of being ``at rest'' is
absolutely defined. Indeed, lattice models of multi-lane traffic flow have
recently been studied \cite{WNW}. The present model could be interpreted as a
multilane flow in which vehicles cannot pass one another without an empty lane
being present between them as they do so. However, the rules adopted for
determining cluster velocities would themselves then have to be chosen
non-Galilean-invariant (a cluster would travel only as fast as its slowest
member). Apart from rotation effects \cite{KB}, the presence or absence of
Galilean invariance is (arguably) the major difference between colloidal and
vehicular jamming.

In summary, the counter scheme \cite{SH} is a deterministic way of implementing
cluster flow which circumvents some of the difficulties associated with a
stochastic implementation of cluster flow. In addition, it only allows single
step movements of clusters and thus avoids potential problems associated with
collisions of clusters and double occupancy of sites.


\section{Discussion}			\label{sec:Conclusion}

In this work we have studied a simple stochastic model of shear-induced
clustering on a $2d$ lattice. Our results strongly suggest that a discontinuous
transition occurs, with strong hysteresis, from an unjammed state to a jammed
state as the shear rate is increased or temperature decreased. As the system
jams, the distribution of clusters becomes power law. However, the exponent may
have a non-trivial temperature dependence, appearing to vary continuously.

Our conclusions are based on the study of finite systems and, while we have
presented strong evidence in favour of the presence of a discontinuous phase
transition in the model, care is required in identifying such a phase
transition \cite{OEC}. It would be useful therefore to study systematically
finite-size effects in the system, as well as the effect of varying the aspect
ratio. Also, we have not considered rotation of clusters but it should be
possible to do so following ref.~\cite{KB}.

It is interesting to compare the present transition to several other models of
reversible cluster-cluster aggregation in which the aggregation is due to
diffusion of clusters, rather than an imposed shear. In the conserved-mass
aggregation model of Krishnamurthy {\it et al} \cite{KMB} a transition from a
disordered (unjammed) phase to a phase containing an infinite cluster occurs as
the rate of single particle dissociation increases. The infinite aggregate
phase exhibits a power law cluster size distribution. However, in contrast to
the present model, the transition in that work is reminiscent of Bose
condensation and does not exhibit hysteresis. We also mention the work of Shih
{\it et al\/} \cite{SLSA} and Haw {\it et al\/} \cite{HSPP} where diffusion
limited cluster-cluster aggregation with finite bond energy (thus allowing
fragmentation) is studied in the context of colloidal gels.

\ack{OJO is supported by a University of Edinburgh Postgraduate Research 
	Studentship and gratefully acknowledges the award of a Royal Society Summer
	Studentship during the initial part of this work. MRE is a Royal Society
	University Research Fellow.}



\begin{thebibliography}{99}

\bibitem{Colloids} see e.g., W.J. Frith, P. D'Haene, R. Buscall and J. Mewis,
		J. Rheol. 40 (1996) 531;
	J.~Bender and N.J.~Wagner, J. Rheol. 40 (1996) 899.
 
\bibitem{Micelle}
	I. Wunderlich, H. Hoffmann and H. Rehage,
		Rheol. Acta 26 (1987) 532;
	R. Bruinsma, W.M. Gelbart and A. Ben-Shaul,
		J. Chem. Phys. 96 (1992) 7710;
	P. Boltenhagen, Y. Hu, E.F. Matthys and D.J. Pine,
		Phys. Rev. Lett. 79 (1997) 2359.

\bibitem{BB} J.F. Brady and G. Bossis, Annu. Rev. Fluid Mech. 20 (1988) 111.

\bibitem{FMB} R.S. Farr, J.R. Melrose and R.C. Ball,
        Phys. Rev. E 55 (1997) 7203.

\bibitem{SZ} Statistical Mechanics of Driven Diffusive Systems,
	eds. B.~Schmittman and R.K.P.~Zia (Academic Press, UK, 1995).

\bibitem{Krug} For a review, see J. Krug, to appear in: 
	Traffic and Granular Flow '97, (Springer, Singapore, 1998).

\bibitem{OEC} O.J.~O'Loan, M.R.~Evans and M.E.~Cates,
	{\it preprint} cond-mat/9712112.

\bibitem{BML} O.Biham, A.A. Middleton and D. Levine,
	Phys. Rev. A 46 (1992) 6124.

\bibitem{SHZ} B. Schmittmann, K. Hwang and R.K.P. Zia,
	Europhys. Lett. 19 (1992) 19.

\bibitem{SH} S.B.~Santra and H.J.~Herrmann,
	Physica A 218 (1995) 298.

\bibitem{KB} T.~Kov\'acs and G.~B\'ardos, Physica A 234 (1997) 665.

\bibitem{VE1} P.G.J. van Dongen and M.H. Ernst,
	J. Stat. Phys. 50 (1988) 295.

\bibitem{VE2} P.G.J. van Dongen and M.H. Ernst, 
	Phys. Rev. Lett. 54 (1985) 1396.

\bibitem{WNW} see e.g., P. Wagner, K. Nagel and D.E. Wolf,
	Physica A 234 (1997) 687.

\bibitem{KMB} S. Krishnamurthy, S.N. Majumdar and M. Barma,
	{\it preprint} cond-mat/9709117.

\bibitem{SLSA} W.Y. Shih, J. Liu, W.H. Shih and I.A. Aksay,
	J. Stat. Phys. 62 (1991) 961.

\bibitem{HSPP} M.D. Haw, M. Sievwright, W.C.K. Poon and P.N. Pusey,
	Adv. Colloid Interface Sci. 62 (1995) 1.

\end{thebibliography}
\end{document}